\documentclass[a4paper,11pt]{article}
\usepackage[margin=1in]{geometry}
\usepackage{amssymb, amsmath, amsthm, mathrsfs}
\usepackage{cite}
\usepackage{tabls}
\usepackage{multirow}
\usepackage{pifont}
\usepackage{graphicx}

\def\qed{\hfill $\Box$\vspace{0.3cm}}
\def\pf{\noindent{\bf Proof. }}

\newtheorem{lemma}{Lemma}
\newtheorem{theorem}{Theorem}

\newtheorem{corollary}[lemma]{Corollary}

\def\pf{\noindent{\bf Proof. }}

\begin{document}

\title{\LARGE\bf The 4-Component Connectivity of Alternating Group Networks\thanks{This research was partially supported by the grant MOST-107-2221-E-141-001-MY3 from the Ministry of Science and Technology, Taiwan.
}
}

\author{
Jou-Ming Chang$^{1,}$\thanks{Corresponding author. Email: spade@ntub.edu.tw} 
\hspace{.1in} Kung-Jui Pai$^{2}$
\hspace{.1in} Ro-Yu Wu$^{3}$
\hspace{.1in} Jinn-Shyong Yang$^{4}$
\\
\\
{\small $^1$ Institute of Information and Decision Sciences,}\\
{\small National Taipei University of Business, Taipei, Taiwan, ROC}\\ 
{\small $^2$ Department of Industrial Engineering and Management,}\\
{\small Ming Chi University of Technology, New Taipei City, Taiwan, ROC} \\ 
{\small $^3$ Department of Industrial Management,}\\
{\small Lunghwa University of Science and Technology, Taoyuan, Taiwan, ROC}\\
{\small $^4$ Department of Information Management,}\\
{\small National Taipei University of Business, Taipei, Taiwan, ROC} \\ 
}
\date{}
\maketitle

\begin{abstract}
\baselineskip=16pt
The $\ell$-component connectivity (or $\ell$-connectivity for short) of a graph $G$, denoted by $\kappa_\ell(G)$, is the minimum number of vertices whose removal from $G$ results in a disconnected graph with at least $\ell$ components or a graph with fewer than $\ell$ vertices. This generalization is a natural extension of the classical connectivity defined in term of minimum vertex-cut. As an application, the $\ell$-connectivity can be used to assess the vulnerability of a graph corresponding to the underlying topology of an interconnection network, and thus is an important issue for reliability and fault tolerance of the network. So far, only a little knowledge of results have been known on $\ell$-connectivity for particular classes of graphs and small $\ell$'s. In a previous work, we studied the $\ell$-connectivity on $n$-dimensional alternating group networks $AN_n$ and obtained the result $\kappa_3(AN_n)=2n-3$ for $n\geqslant 4$. In this sequel, we continue the work and show that $\kappa_4(AN_n)=3n-6$ for $n\geqslant 4$.

\vskip 0.2in 
\noindent 
{\bf Keyword:} Interconnection networks, Graph connectivity, Generalized graph connectivity, Component connectivity, Alternating group networks
\end{abstract}
\setcounter{page}{1}
\baselineskip=16pt

\newpage
\section{Introduction}
\label{sec:intro}

As usual, the underlying topology of an interconnection network is modeled by a connected graph $G=(V,E)$, where $V(=V(G))$ is the set of processors and $E(=E(G))$ is the set of communication links between processors. A subgraph obtained from $G$ by removing a set $F$ of vertices is denoted by $G-F$. A \emph{separating set} (or \emph{vertex-cut}) of a connected graph $G$ is a set $F$ of vertices whose removal renders $G-F$ to become disconnected. If $G$ is not a complete graph, the \emph{connectivity} $\kappa(G)$ is the cardinality of a minimum separating set of $G$. By convention, the connectivity of a complete graph with $n$ vertices is defined to be $n-1$. A graph $G$ is \emph{$n$-connected} if $\kappa(G)\geqslant n$. 

The connectivity is an important topic in graph theory. In particular, it plays a key role in applications related to the modern interconnection networks, e.g., $\kappa(G)$ can be used to assess the vulnerability of the corresponding network, and is an important measurement for reliability and fault tolerance of the network~\cite{Xu-2001}. However, to further analyze the detailed situation of the disconnected network caused by a separating set, it is natural to generalize the classical connectivity by introducing some conditions or restrictions on the separating set $F$ and/or the components of $G-F$~\cite{Harary-1983}. The most basic consideration is the number of components associated with the disconnected network. To figure out what kind of separating sets and/or how many sizes of a separating set can result in a disconnected network with a certain number of components, Chartrand et al.~\cite{Chart-1984} proposed a generalization of connectivity with respect to separating set for making a more thorough study. In this paper, we follow this direction to investigate such kind of generalized connectivity on a class of interconnection networks called alternating group networks (defined later in Section~\ref{sec:Background}).

For an integer $\ell\geqslant 2$, the \emph{generalized $\ell$-connectivity} of a graph $G$, denoted by $\kappa_\ell(G)$, is the minimum number of vertices whose removal from $G$ results in a disconnected graph with at least $\ell$ components or a graph with fewer than $\ell$ vertices. A graph $G$ is \emph{$(n,\ell)$-connected} if $\kappa_\ell(G)\geqslant n$. A synonym for such a generalization was also called the \emph{general connectivity} by Sampathkumar~\cite{Sampathkumar-1984} or \emph{$\ell$-component connectivity} (\emph{$\ell$-connectivity} for short) by Hsu et al.~\cite{Hsu-2012}, Cheng et al.~\cite{Eddie-2014,Eddie-2015,Eddie-2017} and Zhao et al.~\cite{Zhao-2016}. Hereafter, we follow the use of the terminology of Hsu et al. Obviously, $\kappa_2(G)=\kappa(G)$. Similarly, for an integer $\ell\geqslant 2$, the \emph{generalized $\ell$-edge-connectivity} (\emph{$\ell$-edge-connectivity} for short) $\lambda_\ell(G)$, which was introduced by Boesch and Chen~\cite{Boesch-1978}, is defined to be the smallest number of edges whose removal leaves a graph with at least $\ell$ components if $|V(G)|\geqslant\ell$, and $\lambda_\ell(G)=|V(G)|$ if $|V(G)|<\ell$. In addition, many problems related to networks on faulty edges haven been considered in~\cite{Hen-2003,Hsi-2004,Hsi-2001,Pan-2003}.

The notion of $\ell$-connectivity is concerned with the relevance of the cardinality of a minimum vertex-cut and the number of components caused by the vertex-cut, which is a good measure of robustness of interconnection networks. Accordingly, this generalization is called the \emph{cut-version} definition of generalized connectivity. We note that there are other diverse generalizations of connectivity in the literature, e.g., Hager~\cite{Hager-1985} gave the so-called \emph{path-version} definition of generalized connectivity, which is defined from the view point of Menger's Theorem. Recently, Sun and Li~\cite{Sun-2017} gave sharp bounds of the difference between the two versions of generalized connectivities. 

For research results on $\ell$-connectivity of graphs, the reader can refer to~\cite{
Chart-1984,
Day-1991,
Day-1999,
Eddie-2014,
Eddie-2015,
Eddie-2017,
Hsu-2012,
Oellermann-1987-A,
Oellermann-1987-B,
Sampathkumar-1984,
Zhao-2016}. 
At the early stage, the main work focused on establishing sufficient conditions for graphs to be $(n,\ell)$-connected, (e.g., see~\cite{Chart-1984,Sampathkumar-1984,Oellermann-1987-A}). Also, several sharp bounds of $\ell$-connectivity related to other graph parameters can be found in~\cite{Sampathkumar-1984,Day-1999}. In addition, for a graph $G$ and an integer $k\in[0,\kappa_\ell(G)]$, a function called \emph{$\ell$-connectivity function} is defined to be the minimum $\ell$-edge-connectivity among all subgraphs of $G$ obtained by removing $k$ vertices from $G$, and several properties of this function was investigated in~\cite{Day-1991,Oellermann-1987-B}. By contrast, finding  $\ell$-connectivity for certain interconnection networks is a new trend of research at present. So far, the exact values of $\ell$-connectivity are known only for a few classes of networks, in particular, only for small $\ell$'s. For example, $\kappa_\ell(G)$ is determined on the $n$-dimensional hypercube for $\ell\in[2,n+1]$ (see~\cite{Hsu-2012}) and $\ell\in[n+2,2n-4]$ (see~\cite{Zhao-2016}), the $n$-dimensional hierarchical cubic network (see~\cite{Eddie-2014}), the $n$-dimensional complete cubic network (see~\cite{Eddie-2015}), and the generalized exchanged hypercube $GEH(s,t)$ for $1\leqslant s\leqslant t$ and $\ell\in[2,s+1]$ (see~\cite{Eddie-2017}). However, determining $\ell$-connectivity is still unsolved for most interconnection networks. As a matter of fact, it has been pointed out in~\cite{Hsu-2012} that, unlike the hypercube, the results of the well-known interconnection networks such as the star graphs~\cite{Akers-1987} and the alternating group graphs~\cite{Jwo-1993} are still unknown.

Recently, we studied two types of generalized 3-connectivities (i.e., the cut-version and the path-version of the generalized connectivities as mentioned before) in the $n$-dimensional alternating group network $AN_n$, which was introduced by Ji~\cite{Ji-1998} to serve as an interconnection network topology for computing systems. 
In~\cite{Chang-2017}, we already determined the 3-component connectivity $\kappa_3(AN_n)=2n-3$ for $n\geqslant 4$. In this sequel, we continue the work and show the following result.

\begin{theorem} \label{thm:main}
For $n\geqslant 4$, $\kappa_4(AN_n)=3n-6$.
\end{theorem}

\section{Background of alternating group networks}
\label{sec:Background}

Let ${\mathbb Z}_n=\{1,2,\ldots,n\}$ and $A_n$ denote the set of all even permutations over ${\mathbb Z}_n$. For $n\geqslant 3$, the \emph{$n$-dimensional alternating group network}, denoted by $AN_n$, is a graph with the vertex set of even permutations (i.e., $V(AN_n)=A_n$), and two vertices $p=(p_1p_2\cdots p_n)$ and $q=(q_1q_2\cdots q_n)$ are adjacent if and only if one of the following three conditions holds~\cite{Ji-1998}:
\begin{description}
\vspace{-0.4cm}
\item[] (i) $p_1=q_2$, $p_2=q_3$, $p_3=q_1$, and $p_j=q_j$ for $j\in{\mathbb Z}_n\setminus\{1,2,3\}$.
\vspace{-0.2cm}
\item[] (ii) $p_1=q_3$, $p_2=q_1$, $p_3=q_2$, and $p_j=q_j$ for $j\in{\mathbb Z}_n\setminus\{1,2,3\}$.
\vspace{-0.2cm}
\item[] (iii) There exists an $i\in\{4,5,\ldots,n\}$ such that 
$p_1=q_2$, $p_2=q_1$, $p_3=q_i$, $p_i=q_3$, and $p_j=q_j$ for $j\in{\mathbb Z}_n\setminus\{1,2,3,i\}$.
\end{description}

\begin{figure}[ht]
\begin{center}
\includegraphics[width=0.95\textwidth]{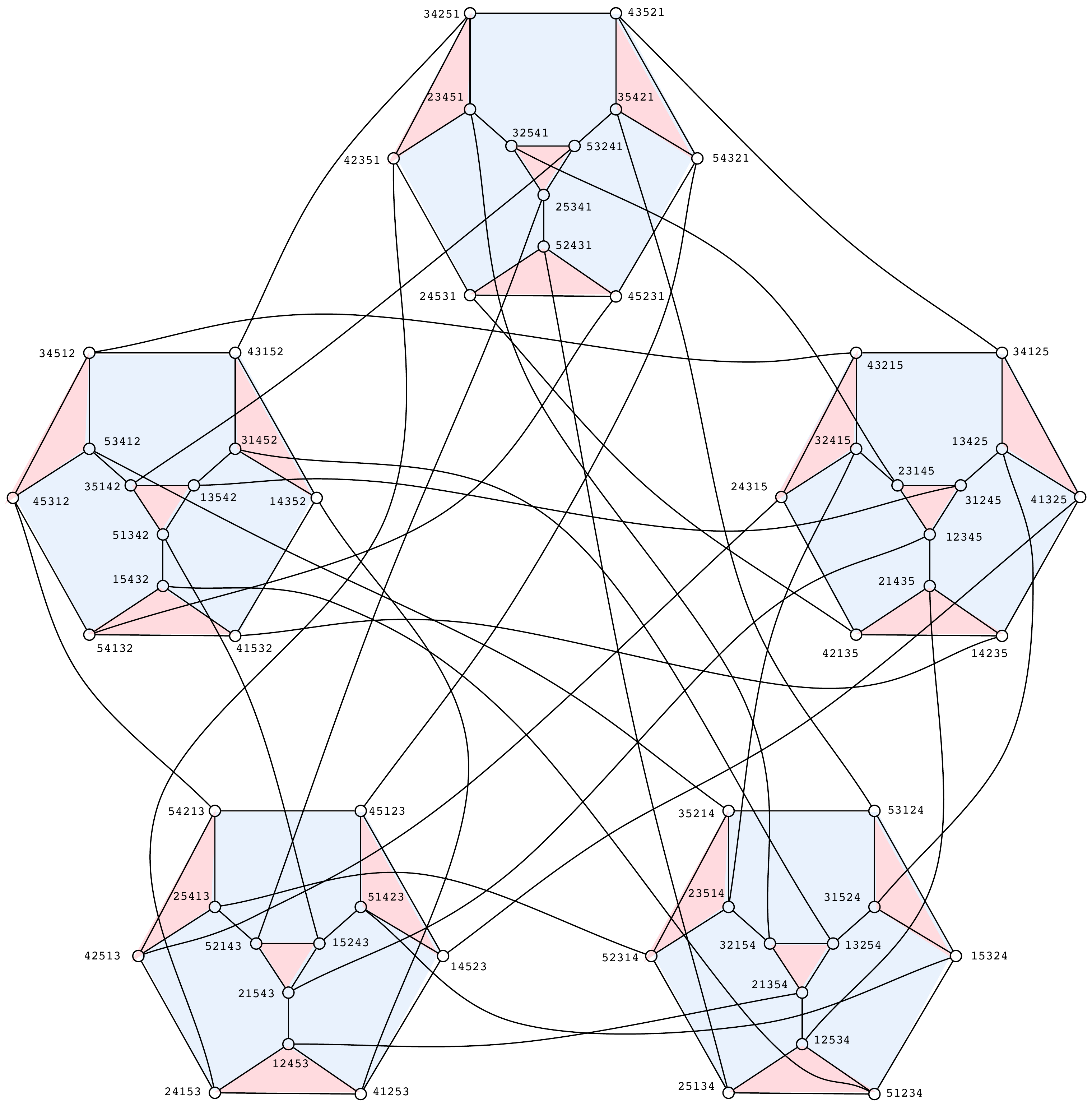}
\caption{Alternating group network $AN_5$.} 
\label{fig:AN5}
\end{center}
\end{figure}

The basic properties of $AN_n$ are known as follows. $AN_n$ contains $n!/2$ vertices and $n!(n-1)/4$ edges, which is a vertex-symmetric and $(n-1)$-regular graph with diameter $\lceil 3n/2\rceil-3$ and connectivity $n-1$. For $n\geqslant 3$ and $i\in{\mathbb Z}_n$, let $AN_n^i$ be the subnetwork of $AN_n$ induced by vertices with the rightmost symbol $i$ in its permutation. It is clear that $AN_n^i$ is isomorphic to $AN_{n-1}$. In fact, $AN_n$ has a recursive structure, which can be constructed from $n$ disjoint copies $AN_n^i$ for $i\in{\mathbb Z}_n$ such that, for any two subnetworks $AN_n^i$ and $AN_n^j$, $i,j\in{\mathbb Z}_n$ and $i\ne j$, there exist $(n-2)!/2$ edges between them. Fig.~\ref{fig:AN5} depicts $AN_5$, where each part of shadows indicates a subnetwork isomorphic to $AN_4$.  

A path (resp., cycle) of length $k$ is called a \emph{$k$-path} (resp., \emph{$k$-cycle}). For notational convenience, if a vertex $x$ belongs to a subnetwork $AN_n^i$, we simply write $x\in AN_n^i$ instead of $x\in V(AN_n^i)$. The disjoint union of two subnetworks $AN_n^i$ and $AN_n^j$ is denoted by $AN_n^i\cup AN_n^j$. The subgraph obtained from $AN_n$ by removing a set $F$ of vertices is denoted by $AN_n-F$. An edge $(x,y)\in E(AN_n)$ with two end vertices $x\in AN_n^i$ and $y\in AN_n^j$ for $i\ne j$ is called an \emph{external edges} between $AN_n^i$ and $AN_n^j$. In this case, $x$ and $y$ are called \emph{out-neighbors} to each other. By contrast, edges joining vertices in the same subnetwork are called \emph{internal edges}, and the two adjacent vertices are called \emph{in-neighbors} to each other. By definition, it is easy to check that every vertex of $AN_n$ has $n-2$ in-neighbors and exactly one out-neighbor. Hereafter, for a vertex $x\in AN_n$, we use $N(x)$ to denote the set of in-neighbors of $x$, and $\text{out}(x)$ the unique out-neighbor of $x$. Moreover, if $H$ is a subgraph of $AN_n^i$, we define $N(H)=(\bigcup_{x\in V(H)} N(x))\setminus V(H)$ as the \emph{in-neighborhood} of $H$, i.e., the set composed of all in-neighbors of those vertices in $H$ except for those belong to $H$.

In what follow, we shall present some properties of $AN_n$, which will be used later. For more properties on alternating group networks, we refer to~\cite{Chen-2006,Hao-2012,Ji-1998,Zhou-2010,Zhou-2009}.

\begin{lemma}\label{lm:basic} {\rm (see~\cite{Hao-2012,Zhou-2010,Zhou-2009})}
For $AN_n$ with $n\geqslant 4$ and $i,j\in{\mathbb Z}_n$ with $i\ne j$, the following holds:
\begin{description}
\vspace{-0.4cm}
\item {\rm(1)} $AN_n$ has no $4$-cycle and $5$-cycle.
\vspace{-0.2cm}
\item {\rm(2)} Any two distinct vertices of $AN_n^i$ have different out-neighbors in $AN_n-V(AN_n^i)$.
\vspace{-0.2cm}
\item {\rm(3)} There are exactly $(n-2)!/2$ edges between $AN_n^i$ and $AN_n^j$.
\end{description}
\end{lemma}

\begin{lemma}\label{lm:subgraphs}
For $n\geqslant 6$ and $i\in{\mathbb Z}_n$, let $H$ be a connected induced subgraph of $AN_n^i$. Then, the following properties hold:
\begin{description}
\vspace{-0.4cm}
\item {\rm(1)} If $|V(H)|=3$, then $H$ is a $3$-cycle or a $2$-path. Moreover, if $H$ is a $3$-cycle {\rm(}resp., a $2$-path{\rm)}, then $|N(H)|=3n-12$ {\rm(}resp., $3n-11\leqslant |N(H)|\leqslant 3n-10${\rm)}.   
\vspace{-0.2cm}
\item {\rm(2)} If $4\leqslant|V(H)|<(n-1)!/4$, then $|N(H)|\geqslant 4n-16$.
\end{description}
\end{lemma}
\pf 
The two properties can easily be proved by induction on $n$. Now, we only verify the subgraph $H$ in Fig.~\ref{fig:AN5} for the basis case $n=6$. Recall that every vertex has $n-2$ in-neighbors in $AN_n^i$. For (1), the result of 3-cycle is clear. If $H$ is a 2-path, at most two adjacent vertices in $H$ can share a common in-neighbor, it follows the $3n-11\leqslant|N(H)|\leqslant 3n-10$. For (2), the condition $|V(H)|<(n-1)!/4$ means that the number of vertices in $H$ cannot exceed a half of those in $AN_n^i$. In particular, if $|V(H)|=4$, then $H$ is either a claw (i.e., $K_{1,3}$), a paw (i.e., $K_{1,3}$ plus an edge), or a $3$-path. Moreover, if $H$ is a paw, a claw or a 3-path, then no two adjacent vertices, at most one pair of adjacent vertices, or at most two pair of adjacent vertices in $H$ can share a common in-neighbor, respectively. This shows that $|N(H)|=4n-16$ when $H$ is a paw, $4n-15\leqslant|N(H)|\leqslant 4n-14$ when $H$ is a claw, and $4n-16\leqslant|N(H)|\leqslant 4n-14$ when $H$ is a 3-path. Also, if $4<|V(H)|<(n-1)!/4$, it is clear that $|N(H)|>4n-16$.
\qed

For designing a reliable probabilistic network, Bauer et al.~\cite{Bauer-1981} first introduced the notion of super connectedness. A regular graph is (\emph{loosely}) \emph{super-connected} if its only minimum vertex-cuts are those induced by the neighbors of a vertex, i.e., a minimum vertex-cut is the set of neighbors of a single vertex. If, in addition, the deletion of a minimum vertex-cut results in a graph with two components and one of which is a singleton, then the graph is \emph{tightly super-connected}. More accurately, a graph is \emph{tightly $k$-super-connected} provided it is tightly super-connected and the cardinality of a minimum vertex-cut is equal to $k$. Zhou and Xiao~\cite{Zhou-2010} pointed out that $AN_3$ and $AN_4$ are not super-connected, and showed that $AN_n$ for $n\geqslant 5$ is tightly $(n-1)$-super-connected. Moreover, to evaluate the size of the connected components of $AN_n$ with a set of faulty vertices, Zhou and Xiao gave the following properties.

\begin{lemma}\label{lm:two-components} {\rm (see~\cite{Zhou-2010})}
For $n\geqslant 5$, if $F$ is a vertex-cut of $AN_n$ with $|F|\leqslant 2n-5$, then one of the following conditions holds:
\begin{description}
\vspace{-0.4cm}
\item {\rm(1)} $AN_n-F$ has two components, one of which is a trivial component {\rm(}i.e., a singleton{\rm)}.
\vspace{-0.2cm}
\item {\rm(2)} $AN_n-F$ has two components, one of which is an edge, say $(u,v)$. In particular, if $|F|=2n-5$, $F$ is composed of all neighbors of $u$ and $v$, excluding $u$ and $v$.
\end{description}
\end{lemma}

\begin{lemma}\label{lm:three-components} {\rm (see~\cite{Zhou-2010})}
For $n\geqslant 5$, if $F$ is a vertex-cut of $AN_n$ with $|F|\leqslant 3n-10$, then one of the following conditions holds:
\begin{description}
\vspace{-0.4cm}
\item {\rm(1)} $AN_n-F$ has two components, one of which is either a singleton or an edge.
\vspace{-0.2cm}
\item {\rm(2)} $AN_n-F$ has three components, two of which are singletons.
\end{description}
\end{lemma}

Through a more detailed analysis, Chang et al.~\cite{Chang-2017} recently obtained a slight extension of the result of Lemma~\ref{lm:two-components} as follows.

\begin{lemma}\label{lm:two-components-improved} {\rm (see~\cite{Chang-2017})}
Let $F$ is a vertex-cut of $AN_n$ with $|F|\leqslant 2n-4$. Then, the following conditions hold:
\begin{description}
\vspace{-0.4cm}
\item {\rm(1)} If $n=4$, then $AN_n-F$ has two components, one of which is a singleton, an edge, a $3$-cycle, a $2$-path, or a paw.
\vspace{-0.2cm}
\item {\rm(2)} If $n=5$, then $AN_n-F$ has two components, one of which is a singleton, an edge, or a $3$-cycle.
\vspace{-0.2cm}
\item {\rm(3)} 
If $n\geqslant 6$, then $AN_n-F$ has two components, one of which is either a singleton or an edge.
\end{description}
\end{lemma}

\section{The 4-component connectivity of $AN_n$}
\label{sec:Cut}

Since $AN_3$ is a 3-cycle, by definition, it is clear that $\kappa_4(AN_3)=1$. Also, in the process of the drawing of Fig.~\ref{fig:AN5}, we found by a brute-force checking that the removal of no more than five vertices in $AN_4$ (resp., eight vertices in $AN_5$) results in a graph that is either connected or contains at most three components. Thus, the following lemma establishes the lower bound of $\kappa_4(AN_n)$ for $n=4,5$.

\begin{lemma}\label{lm:n4n5}
$\kappa_4(AN_4)\geqslant 6$ and $\kappa_4(AN_5)\geqslant 9$.
\end{lemma}

\begin{lemma}\label{lm:lower-bound}
For $n\geqslant 6$, $\kappa_4(AN_n)\geqslant 3n-6$.
\end{lemma}
\pf
Let $F$ be any vertex-cut in $AN_n$ such that $|F|\leqslant 3n-7$. For convenience, vertices in $F$ (resp., not in $F$) are called faulty vertices (resp., fault-free vertices). By Lemma~\ref{lm:three-components}, if $|F|\leqslant 3n-10$, then $AN_n-F$ contains at most three components. To complete the proof, we need to show that the same result holds for $3n-9\leqslant|F|\leqslant 3n-7$. Let $F_i=F\cap V(AN_n^i)$ and $f_i=|F_i|$ for each $i\in{\mathbb Z}_n$. We claim that there exists some subnetwork, say $AN_n^i$, such that it contains $f_i\geqslant n-2$ faulty vertices. Since $3(n-2)>3n-7\geqslant|F|$, if it is so, then there are at most two such subnetworks. Suppose not, i.e., every subnetwork $AN_n^j$ for $j\in{\mathbb Z}_n$ has $f_j\leqslant n-3$ faulty vertices. Since $AN_n^j$ is $(n-2)$-connected, $AN_n^j-F_j$ remains connected for each $j\in{\mathbb Z}_n$. Recall the property~(3) of Lemma~\ref{lm:basic} that there are $(n-2)!/2$ independent edges between $AN_n^i$ and $AN_n^j$ for each pair $i,j\in {\mathbb Z}_n$ with $i\ne j$. Since $(n-2)!/2>2(n-3)\geqslant f_i+f_j$ for $n\geqslant 6$, it guarantees that the two subgraphs $AN_n^i-F_i$ and $AN_n^j-F_j$ are connected by an external edge in $AN_n-F$. Thus, $AN_n-F$ is connected, and this contradicts to the fact that $F$ is a vertex-cut in $AN_n$. Moreover, for such subnetworks, it is sure that some of $F_i$ must be a vertex-cut of $AN_n^i$. Otherwise, $AN_n-F$ is connected, a contradiction. We now consider the following two cases:

{\bf Case~1}: There is exactly one such subnetwork, say $AN_n^i$, such that it contains $f_i\geqslant n-2$ faulty vertices. In this case, we have $f_j\leqslant n-3$ for all $j\in{\mathbb Z}_n\setminus\{i\}$ and $F_i$ is a vertex-cut of $AN_n^i$. Let $H$ be the subgraph of $AN_n$ induced by the fault-free vertices outside $AN_n^i$, i.e., $H=AN_n-(V(AN_n^i)\cup F)$. Since every subnetwork $AN_n^j$ in $H$ has $f_j\leqslant n-3$ faulty vertices, from the previous argument it is sure that $H$ is connected. We denote by $C$ the component of $AN_n-F$ that contains $H$ as its subgraph, and let $f=|F|-f_i$ be the number of faulty vertices outside $AN_n^i$. Since $3n-7\geqslant|F|\geqslant f_i\geqslant n-2$, we have $0\leqslant f\leqslant 2n-5$. Consider the following scenarios:

{\bf Case~1.1}: $f=0$. In this case, there are no faulty vertices outside $AN_n^i$. That is, $H=AN_n-V(AN_n^i)$. Indeed, this case is impossible because if it is the case, then every vertex of $AN_n^i-F_i$ has the fault-free out-neighbor in $H$. Thus, $AN_n^i-F_i$ belongs to $C$, and it follows that $AN_n-F$ is connected, a contradiction. 

{\bf Case~1.2}: $f=1$. Let $u\in F\setminus F_i$ be the unique faulty vertex outside $AN_n^i$. That is, $H=AN_n-(V(AN_n^i)\cup\{u\})$. Since $F_i$ is a vertex-cut of $AN_n^i$, we assume that $AN_n^i-F_i$ is divided into $k$ disjoint connected components, say $C_1,C_2,\ldots, C_k$. For each $j\in{\mathbb Z}_k$, if $|C_j|\geqslant 2$, then there is at least one vertex of $C_j$ with its out-neighbor in $H$, and thus $C_j$ belongs to $C$. We now consider a component that is a singleton, say $C_j=\{v\}$. If $\text{out}(v)\ne u$, then $\text{out}(v)$ must be contained in $H$, and thus $C_j$ belongs to $C$. Clearly, there exists at most one component $C_j=\{v\}$ such that $\text{out}(v)=u$. In this case, $AN_n-F$ has exactly two components $\{v\}$ and $C$. 

{\bf Case~1.3}: $f=2$. Let $u_1,u_2\in F\setminus F_i$ be the two faulty vertices outside $AN_n^i$. That is, $H=AN_n-(V(AN_n^i)\cup\{u_1,u_2\})$. Since $F_i$ is a vertex-cut of $AN_n^i$, we assume that $AN_n^i-F_i$ is divided into $k$ disjoint connected components, say $C_1,C_2,\ldots, C_k$. For each $j\in{\mathbb Z}_k$, if $|C_j|\geqslant 3$, then there is at least one vertex of $C_j$ with its out-neighbor in $H$, and thus $C_j$ belongs to $C$. We now consider a component $C_j$ with $|C_j|=2$, i.e., $C_j$ is an edge, say $(v,w)$. By the property~(2) of Lemma~\ref{lm:basic}, we have $\text{out}(v)\ne\text{out}(w)$. If $\{\text{out}(v),\text{out}(w)\}\ne\{u_1,u_2\}$, then at least one of $\text{out}(v)$ and $\text{out}(w)$ must be contained in $H$, and thus $C_j$ belongs to $C$. Since $(3n-7)-2\geqslant f_i=|F|-f\geqslant (3n-9)-2$ and $(v,w)$ has $2n-6$ in-neighbors (not including $v$ and $w$) in $AN_n^i$, we have $2n-6<f_i<2(2n-6)$ for $n\geqslant 6$. Thus, there exists at most one such component $C_j=\{(v,w)\}$ such that $\{\text{out}(v),\text{out}(w)\}=\{u_1,u_2\}$. If it is the case of existence, then $AN_n-F$ has exactly two components $\{(v,w)\}$ and $C$. Finally, we consider a component that is a singleton. Since $3n-9\leqslant f_i\leqslant 3n-11$ and every vertex has degree $n-2$ in $AN_n^i$, we have $n-2<f_i<3(n-2)$ for $n\geqslant 6$. Thus, at most two such components exist in $AN_n^i-F_i$, say $C_j=\{v\}$ and $C_{j'}=\{w\}$ where $j,j'\in{\mathbb Z}_k$. If $\text{out}(v),\text{out}(w)\notin\{u_1,u_2\}$, then both $\text{out}(v)$ and $\text{out}(w)$ must be contained in $H$, and thus $C_j$ and $C_{j'}$ belong to $C$. Also, if either $\text{out}(v)\notin\{u_1,u_2\}$ or $\text{out}(w)\notin\{u_1,u_2\}$, then $AN_n-F$ has exactly two components, one of which is a singleton $\{v\}$ or $\{w\}$. Finally, if $\{\text{out}(v),\text{out}(w)\}=\{u_1,u_2\}$, then $AN_n-F$ has exactly three components, two of which are singletons $\{v\}$ and $\{w\}$.

{\bf Case~1.4}: $f=3$. Let $u_1,u_2,u_3\in F\setminus F_i$ be the three faulty vertices outside $AN_n^i$. That is, $H=AN_n-(V(AN_n^i)\cup\{u_1,u_2,u_3\})$. Since $F_i$ is a vertex-cut of $AN_n^i$, we assume that $AN_n^i-F_i$ is divided into $k$ disjoint connected components, say $C_1,C_2,\ldots, C_k$. For each $j\in{\mathbb Z}_k$, if $|C_j|\geqslant 4$, then there is at least one vertex of $C_j$ with its out-neighbor in $H$, and thus $C_j$ belongs to $C$. We now consider a component $C_j$ with $|C_j|=3$, i.e., $C_j$ is either a 3-cycle or a 2-path. Assume that $V(C_j)=\{v_1,v_2,v_3\}$. If there is a vertex $\text{out}(v_h)\notin\{u_1,u_2,u_3\}$ for $1\leqslant h\leqslant 3$, then $\text{out}(v_h)$ must be contained in $H$, and thus $C_j$ belongs to $C$. Since $(3n-7)-3\geqslant f_i=|F|-f\geqslant (3n-9)-3$ and, by Lemma~\ref{lm:subgraphs}, we have $3n-12\leqslant |N(C_j)|\leqslant n-10$, it follows that there exists at most one such component $C_j$ such that $\{\text{out}(v_1),\text{out}(v_2),\text{out}(v_3)\}=\{u_1,u_2,u_3\}$. If it is the case of existence, then $AN_n-F$ has exactly two components, one of which is either a 3-cycle or a 2-path. Next, we consider a component $C_j$ with $|C_j|=2$, i.e., $C_j$ is an edge, say $(v,w)$. From an argument similar to Case~1.3 for analyzing the membership of $\text{out}(v)$ and $\text{out}(w)$ in the set $\{u_1,u_2,u_3\}$, we can show that $AN_n-F$ has exactly two components $\{(v,w)\}$ and $C$. Finally, we consider a component that is a singleton. Then, an argument similar to Case~1.3 for analyzing singleton components shows that at most two such components exist in $AN_n^i-F_i$. Thus, $AN_n-F$ has either two components (where one of which is a singleton) or three components (where two of which are singletons).

{\bf Case~1.5}: $f=4$. Let $u_1,u_2,u_3,u_4\in F\setminus F_i$ be the four faulty vertices outside $AN_n^i$. That is, $H=AN_n-(V(AN_n^i)\cup\{u_1,u_2,u_3,u_4\})$. Since $F_i$ is a vertex-cut of $AN_n^i$, we assume that $AN_n^i-F_i$ is divided into $k$ disjoint connected components, say $C_1,C_2,\ldots, C_k$. For each $j\in{\mathbb Z}_k$, if $|C_j|\geqslant 5$, then there is at least one vertex of $C_j$ with its out-neighbor in $H$, and thus $C_j$ belongs to $C$. If $|C_j|\geqslant 4$, by Lemma~\ref{lm:subgraphs}, we have $|N(C_j)|\geqslant 4n-16$. Since $(3n-7)-4\geqslant |F|-f=f_i$, it follows that $|N(C_j)|\geqslant f_i$ for $n\geqslant 6$. Thus, none of component $C_j$ with $|C_j|=4$ exists in $AN_n^i$. Next, we consider a component $C_j$ with $|C_j|=3$ and assume $V(C_j)=\{v_1,v_2,v_3\}$. By Lemma~\ref{lm:subgraphs}, we have $3n-12\leqslant |N(C_j)|\leqslant n-10$. Since $f_i$ is no more than $3n-11$, at most one such component $C_j$ exists in $AN_n^i-F_i$. Furthermore, if such $C_j$ exists, then it is either a 3-cycle or a 2-path. Thus, an argument similar to Case~1.4 for analyzing the membership of $\text{out}(v_1)$, $\text{out}(v_2)$ and $\text{out}(v_3)$ in the set $\{u_1,u_2,u_3,u_4\}$, we can show that $AN_n-F$ has exactly two components, one of which is a 3-cycle or a 2-path. Finally, if we consider a component $C_j$ with $|C_j|\leqslant 2$, an argument similar to the previous cases shows that $AN_n-F$ has either two components (where one of which is a singleton or an edge) or three components (where two of which are singletons).

{\bf Case~1.6}: $f=5$. Let $u_1,u_2,u_3,u_4,u_5\in F\setminus F_i$ be the five faulty vertices outside $AN_n^i$. That is, $H=AN_n-(V(AN_n^i)\cup\{u_1,u_2,u_3,u_4,u_5\})$. Since $F_i$ is a vertex-cut of $AN_n^i$, we assume that $AN_n^i-F_i$ is divided into $k$ disjoint connected components, say $C_1,C_2,\ldots, C_k$. For each $j\in{\mathbb Z}_k$, if $|C_j|\geqslant 6$, then there is at least one vertex of $C_j$ with its out-neighbor in $H$, and thus $C_j$ belongs to $C$. If $|C_j|=4$ or $|C_j|=5$, by Lemma~\ref{lm:subgraphs}, we have $|N(C_j)|\geqslant 4n-16$. Since $(3n-7)-5\geqslant |F|-f=f_i$, it follows that $|N(C_j)|\geqslant f_i$ for $n\geqslant 6$. Thus, none of component $C_j$ with $|C_j|=4$ or $|C_j|=5$ exists in $AN_n^i$. We now consider a component $C_j$ with $|C_j|=3$. Since $f_i\leqslant 3n-12$, by Lemma~\ref{lm:subgraphs}, if such $C_j$ exists, then it must be a 3-cycle, and thus an argument similar to the previous cases shows that $AN_n-F$ has exactly two components, one of which is a 3-cycle. Finally, if we consider a component $C_j$ with $|C_j|\leqslant 2$, an argument similar to the previous cases shows that $AN_n-F$ has either two components (where one of which is a singleton or an edge) or three components (where two of which are singletons). 

{\bf Case~1.7}: $6\leqslant f\leqslant 2n-5$. In this case, we have $(3n-7)-6\geqslant f_i=|F|-f\geqslant (3n-9)-(2n-5)$. Since $AN_n^i$ is isomorphic to $AN_{n-1}$ and $F_i$ is a vertex-cut of $AN_n^i$ with no more than $3(n-1)-10$ vertices, by Lemma~\ref{lm:three-components}, $AN_n^i-F_i$ has at most three components as follows:

{\bf Case~1.7.1}: $AN_n^i-F_i$ has two components, one of which is either a singleton or an edge. Let $C_1$ and $C_2$ be such two components for which $1\leqslant|C_1|\leqslant 2<|C_2|$. More precisely, $|C_2|=|V(AN_n^i)|-f_i-|C_1|\geqslant (n-1)!/2-f_i-2>(3n-7)-f_i\geqslant |F|-f_i=f$ for $n\geqslant 6$. Clearly, the above inequality indicates that there exist some vertices of $C_2$ such that their out-neighbors are contained in $H$, even if all out-neighbors of vertices in $F\setminus F_i$ are contained in $C_2$. Thus, $C_2$ belongs to $C$. Also, if there is a vertex $v\in C_1$ with its out-neighbor in $H$, then $C_1$ belongs to $C$. Otherwise, $AN_n-F$ has exactly two components, one of which is either a singleton or an edge. 

{\bf Case~1.7.2}: $AN_n^i-F_i$ has three components, two of which are singletons. Let $C_1, C_2$ and $C_3$ be such three components for which $|C_1|=|C_2|=1$ and $|C_3|>2$. Since $|C_3|=(n-1)!/2-f_i-2>(3n-7)-f_i\geqslant |F|-f_i=f$ for $n\geqslant 6$, there exist some vertices of $C_2$ such that their out-neighbors are contained in $H$. This shows that $C_2$ belongs to $C$. Since $AN_n^i-F_i$ has three components, the out-neighbor of a vertex $v\in C_1$ or $v\in C_2$ cannot be contained in $H$. Thus, $AN_n-F$ has exactly three components, two of which are singletons. 

{\bf Case~2}: There exist exactly two subnetworks, say $AN_n^i$ and $AN_n^j$, such that $f_i,f_j\geqslant n-2$. Since $F$ is a vertex-cut of $AN_n$, at least one of the subgraphs $AN_n^i-F_i$ and $AN_n^j-F_j$ must be disconnected. Let $H$ be the subgraph of $AN_n$ induced by the fault-free vertices outside $AN_n^i\cup AN_n^j$, i.e., $H=AN_n-(V(AN_n^i)\cup V(AN_n^j)\cup F)$. Since $2n-4\leqslant f_i+f_j\leqslant|F|\leqslant 3n-7$, we have $f_h\leqslant|F|-f_i-f_j\leqslant (3n-7)-(2n-4)=n-3$ for all $h\in{\mathbb Z}_n\setminus\{i,j\}$. The bound of $f_h$ implies that $AN_n^h-F_h$ is connected, and it follows that $H$ is also connected. We denote by $C$ the component of $AN_n-F$ that contains $H$ as its subgraph. Since $n-2\leqslant f_i\leqslant (3n-7)-f_j\leqslant(3n-7)-(n-2)=2n-5$, we consider the following scenarios:

{\bf Case~2.1}: $f_i=2n-5$. Clearly, $f_j\leqslant(3n-7)-f_i=n-2$. Since we have assumed $f_j\geqslant n-2$, it follows that $f_j=n-2$ and there exist no faulty vertices outside $AN_n^i\cup AN_n^j$. That is, $H=AN_n-(V(AN_n^i)\cup V(AN_n^j))$. Indeed, this case is impossible because if it is the case, then there exist a vertex of $(AN_n^i\cup AN_n^j)-F$ such that its out-neighbor is contained in $H$. Thus, $(AN_n^i\cup AN_n^j)-F$ belongs to $C$, and it follows that $AN_n-F$ is connected, a contradiction. 

{\bf Case~2.2}: $n-1\leqslant f_i\leqslant 2n-6$. Since $f_i+f_j\leqslant|F|\leqslant 3n-7$, it implies $f_j\leqslant (3n-7)-f_i\leqslant (3n-7)-(n-1)=2n-6$. Since $AN_n^i$ is isomorphic to $AN_{n-1}$ and $f_i\leqslant 2(n-1)-4$, by Lemma~\ref{lm:two-components-improved}, if $AN_n^i-F_i$ is disconnected, then it has exactly two component, one of which is either a singleton or an edge. Suppose $AN_n^i-F_i=C_1\cup C_2$, where $C_1$ and $C_2$ are disjoint connected components such that $1\leqslant|C_1|\leqslant 2<|C_2|$. More precisely, $|C_2|=|V(AN_n^i)|-f_i-|C_1|=(n-1)!/2-f_i-2>(3n-7)-f_i\geqslant|F|-f_i$ for $n\geqslant 6$, where the last term $|F|-f_i$ is the number of faulty vertices outside $AN_n^i$. Clearly, the above inequality indicates that there exist some vertices of $C_2$ such that their out-neighbors are contained in $H$, even if all out-neighbors of vertices in $F\setminus F_i$ are contained in $C_2$. Thus, $C_2$ belongs to $C$. Also, if there is a vertex of $C_1$ with its out-neighbor in $H$, then $C_1$ belongs to $C$. By contrast, we can show that $AN_n^i-F_i$ belongs to $C$ by a similar way if it is connected. Thus, $AN_n-F$ contains at most one component (which is either a singleton or an edge) such that this component is a subgraph of $AN_n^i$. Similarly, since $f_j\leqslant 2n-6$, $AN_n-F$ contains at most one component (which is either a singleton or an edge) such that this component is a subgraph of $AN_n^j$. Thus, there are at most three components in $AN_n-F$. We claim that $AN_n-F$ cannot simultaneously contain both an edge $(u,v)$ and a singleton $w$ as components. Suppose not and, without loss of generality, assume $u,v\in AN_n^i$ and $w\in AN_n^j$. Then, at least two out-neighbors of $u,v$ and $w$ are not contained in $N(u)\cup N(v)\cup N(w)$. Otherwise, $AN_n$ produces a 4-cycle or 5-cycle, which contradicts to the property~(1) of Lemma~\ref{lm:basic}. Thus, the number of faulty vertices of $AN_n$ requires at least $(2n-6)+(n-2)+2=3n-6\geqslant|F|$, a contradiction. Similarly, we claim that $AN_n-F$ cannot simultaneously contain two disjoint edges $(u_1,v_1)$ and $(u_2,v_2)$ as components. Suppose not. By an argument similar above, we can show that either $AN_n$ has $2(2n-6)+2>3n-7\geqslant|F|$ faulty vertices for $n\geqslant 6$ or it contains a 4-cycle or 5-cycle. However, both the cases are not impossible. Consequently, if $AN_n-F$ contains three component, then two of which are singletons, one is a vertex of $AN_n^i$  and the other is of $AN_n^j$.

{\bf Case~2.3}: $f_i=n-2$. Clearly, $f_j\leqslant(3n-7)-f_i=2n-5$. Since $AN_n^i$ is isomorphic to $AN_{n-1}$ and $n\geqslant 6$, it is tightly $(n-2)$-super-connected. Also, since $f_i=n-2$, if $F_i$ is a vertex-cut of $AN_n^i$, then it must be a minimum vertex-cut. Particularly, there are two components in $AN_n^i-F_i$, one of which is a singleton, say $v$. That is, all in-neighbors of $v$ are faulty vertices (i.e., $N(v)=F_i$). Otherwise, $AN_n^i-F_i$ is connected and thus belongs to $C$. On the other hand, we consider all situations of $AN_n^j-F_j$ as follows. Clearly, if $AN_n^j-F_j$ is connected, then it belongs to $C$, and this further implies that $AN_n^i-F_i$ must be disconnected. In this case, $AN_n-F$ contains exactly two components, one of which is a singleton $v$. We now consider the case that $AN_n^j-F_j$ is not connected and claim that it has at most two disjoint connected components. Suppose not. Since $AN_n^j$ is isomorphic to $AN_{n-1}$, by Lemma~\ref{lm:two-components-improved}, the number of faulty vertices in $AN_n^j$ is at least $2(n-1)-3$. Since $f_j\leqslant 2n-5$, it follows that $f_j=2n-5$. Thus, this situation is a symmetry of Case~2.1 by considering the exchange of $f_i$ and $f_j$, which leads to a contradiction. Suppose $AN_n^j-F_j=C_1\cup C_2$, where $C_1$ and $C_2$ are disjoint connected components such that $|C_1|\leqslant|C_2|$. Since $|C_2|\geqslant (|V(AN_n^j)|-f_j)/2>(n-1)!/4-f_j>(3n-7)-f_j\geqslant|F|-f_j$ for $n\geqslant 6$, where the last term $|F|-f_j$ is the number of faulty vertices outside $AN_n^j$. Clearly, the above inequality indicates that there exist some vertices of $C_2$ such that their out-neighbors are contained in $H$, even if all out-neighbors of vertices in $F\setminus F_j$ are contained in $C_2$. Thus, $C_2$ belongs to $C$. Also, if there is a vertex of $C_1$ with its out-neighbor in $H$, then $C_1$ belongs to $C$. Otherwise, $C_1$ is a component of $AN_n-F$. By Lemma~\ref{lm:subgraphs}, since $f_j=2n-5<4n-16$ when $n\geqslant 6$, we have $|C_1|<4$. Moreover, since $2n-5\leqslant 3n-11$ when $n\geqslant 6$, if $|C_1|=3$, then $C_1$ must be a 3-cycle. If $|C_1|\leqslant 2$, then $C_1$ is either a singleton or an edge. Note that if $C_1$ is a 3-cycle or an edge, then $AN_n-F$ cannot contain the the singleton $v\in V(AN_n^i-F_i)$ as its component. Otherwise, an argument similar to Case~2.2 shows that $AN_n$ either has more than $3n-7$ faulty vertices or produces a 4-cycle or 5-cycle, a contradiction.
\qed

From the proof of Lemma~\ref{lm:lower-bound}, we obtain the following result, which is an extension of Lemma~\ref{lm:three-components}.

\begin{corollary}
For $n\geqslant 5$, if $F$ is a vertex-cut of $AN_n$ with $|F|\leqslant 3n-7$, then one of the following conditions holds:
\begin{description}
\vspace{-0.4cm}
\item {\rm(1)} $AN_n-F$ has two components, one of which is either a singleton, an edge, a $3$-cycle, or a $2$-path.
\vspace{-0.2cm}
\item {\rm(2)} $AN_n-F$ has three components, two of which are singletons.
\end{description}
\end{corollary}

\noindent
{\bf Proof of Theorem~\ref{thm:main}.} 
Lemmas~\ref{lm:n4n5} and \ref{lm:lower-bound} show that $\kappa_4(AN_n)\geqslant 3n-6$ for $n\geqslant 4$. To complete the proof, we need to show the upper bound $\kappa_4(AN_n)\leqslant 3n-6$ for $n\geqslant 4$. Consider an induced 6-cycle $H=(v_1,v_2,v_3,v_4,v_5,v_6)$ in $AN_n$ (the existence of such a cycle can be verified in Fig~\ref{fig:AN5}). Let $F$ be the set composed of all neighbors of vertices in $\{v_1,v_3,v_5\}$. Since every vertex of $AN_n$ has $n-1$ neighbors and any two vertices in $\{v_1,v_3,v_5\}$ share a common neighbor, it is clear that $|F|=3(n-1)-3=3n-6$. Then, the removal of $F$ from $AN_n$ leads to the surviving graph with a large connected component and three singletons $v_1,v_3$ and $v_5$. 
\qed

\section{Concluding remarks}
\label{sec:Conclusion}

In this paper, we follow a previous work to investigate a measure of network reliability, called $\ell$-component connectivity, in alternating group networks $AN_n$. Although we have known that $\kappa_3(AN_n)=2n-3$ and $\kappa_4(AN_n)=3n-6$, at this stage it remains open to determine $\kappa_\ell(AN_n)$ for $\ell\geqslant 5$. Also, as aforementioned, by now little work has been done in determining the $\ell$-component connectivity for most interconnection networks, even if for smaller integer $\ell$. In the future work, we would like to study the $\ell$-component connectivity of $AN_n$ with larger $\ell$, or some popular interconnection networks such as star graphs (and their super class of graphs called $(n,k)$-star graphs and arrangement graphs), bubble-sort graphs, and alternating group graphs.

\end{document}